\documentclass[a4paper, amsfonts, amssymb, amsmath, reprint, showkeys, nofootinbib, twoside, superscriptaddress]{revtex4-1}
\usepackage[english]{babel}
\usepackage[utf8]{inputenc}
\usepackage[colorinlistoftodos, color=green!40, prependcaption]{todonotes}

\usepackage{amsthm}
\usepackage{mathtools}
\usepackage{physics}
\usepackage{xcolor}
\usepackage{graphicx}
\usepackage[left=23mm,right=13mm,top=35mm,columnsep=15pt]{geometry} 
\usepackage{adjustbox}
\usepackage{placeins}
\usepackage[T1]{fontenc}
\usepackage{lipsum}
\usepackage{csquotes}

\usepackage[pdftex, pdftitle={Article}, pdfauthor={Author}]{hyperref} 
\hypersetup{
    colorlinks=true,
    linkcolor=blue,
    filecolor=magenta,      
    urlcolor=cyan,
}

\begin{document}
\title{Supervised and Unsupervised Machine Learning of Structural Phases\\ of Polymers Adsorbed to Nanowires}

\author{Quinn Parker}
    \thanks{Current address: School of Physics, Georgia Institute of Technology, Atlanta, GA 30332, U.S.A.}
    \email[Correspondence email address: ]{qparker3@gatech.edu}
    \affiliation{Department of Physics and Astronomy, University of North Georgia, Dahlonega, GA 30597, U.S.A.}
\author{Dilina Perera}
    \affiliation{Department of Physics, University of Colombo, Colombo 03, Sri Lanka}
\author{Ying Wai Li}
    \affiliation{Computer, Computational, and Statistical Sciences Division, Los Alamos National Laboratory, Los Alamos, NM 87545, U.S.A.}
\author{Thomas Vogel}
        \affiliation{Department of Physics and Astronomy, University of North Georgia, Dahlonega, GA 30597, U.S.A.}
    
\date{\today} 

\begin{abstract}

We identify configurational phases and structural transitions in a polymer nanotube composite by means of machine learning. We employ various unsupervised dimensionality reduction methods, conventional neural networks, as well as the confusion method, an unsupervised neural-network-based approach. We find neural networks are able to reliably recognize all configurational phases that have been found previously in experiment and simulation. Furthermore, we locate the boundaries between configurational phases in a way that removes human intuition or bias. This could be done before only by relying on preconceived, \textit{ad-hoc} order parameters.

\end{abstract}

\maketitle

\section{Introduction} \label{sec:intro}

    We previously studied soft--solid matter nano-com\-posites, in particular systems composed of flexible polymers adsorbed at thin nanostrings or tubes~\cite{Vogel2010prl,Vogel2010pp,Vogel2011cpc}. Such systems are believed to play an important role in current and future development of high-performance nanomaterials. Carbon nanotubes, for example, have been functionalized by wrapping them with certain types of polymers to serve as biosensors for the detection of glucose~\cite{Gao2003ea,Wang2005ea}. However, the successful fabrication of such materials depends on a variety of parameters. In particular, the wetting behavior of carbon nanotubes has been shown to be one critical parameter for the development of nanocomposites~\cite{Tran2008nl}. We have previously developed and employed a coarse-grained model to investigate nanoscale wetting and adhesion phenomena using Monte Carlo methods; we identified various, structurally different low-temperature phases including globular polymers simply attached to the nanostring and polymers completely wrapping, or coating, the substrate~\cite{Vogel2010prl}. One particular problem that we recognized was the classification of structural phases at low temperatures depending on various model parameters. We have addressed the problem in the past by the \textit{ad hoc} introduction of order parameters to identify boundaries between such structural phases. In this paper we show how a less biased approach, based on machine learning, can be deployed. We revisit the earlier introduced configurational phase diagram, aiming at identifying classes of the polymer--wire system and the boundaries between them without any assumptions or other input based on a human perception of structure. 
    In a more general context, such automated structure identification can also provide means to recognize system configurations during Monte Carlo sampling. This could prove beneficial in order to run generalized-ensemble simulations where different structural phases are assigned individual weights~\cite{Schnabel2011jcp}, or simply to collect statistics for individual phases during a simulation. 
    
       Recent years have witnessed significant advances in the use of machine learning (ML) methods for phase classification. In this regard, supervised learning approaches~\cite{Carrasquilla2017np, Schindler2017prb, zhang2017prl, Chng2017prx, Ponte2017prb, Li2018Ann, ZhangP2018prl}, for which the prior labeling of configurations is required, as well as unsupervised learning approaches~\cite{Wang2016prb, Wang2017prb, Hu2017pre, Costa2017prb, Wetzel2017pre, Chng2018pre}, which work without such prior labeling, have been attempted. It has been demonstrated, for example, that neural networks (NNs) trained with labeled configurations can encode information about the ordered and disordered phases in model systems by learning the relevant order parameters~\cite{Carrasquilla2017np}. In particular, approaches based on purposefully mislabeling configurations and evaluating the network performance have been developed to detect phase transitions~\cite{Nieuwenburg2017np}. Such a method does not require true labels to be known in advance and therefore no prior knowledge about the existence (or lack thereof) of a transition is needed. It has also been demonstrated to work in the presence of multiple transitions~\cite{Beach2018prb}.
    
    In the context of unsupervised learning approaches, dimensionality reduction techniques, for example, principal component analysis (PCA), multidimensional scaling (MDS), $t$-distributed stochastic neighbor embedding ($t$-SNE), autoencoders, etc., have been found useful in distinguishing ordered and disordered phases~\cite{Wang2016prb, Wang2017prb, Hu2017pre, Costa2017prb, Wetzel2017pre, Samarakoon2020natcomm}. For systems with clear order parameters, such as the Ising model or the XY model, the latent parameters or the dominant principal components have been shown to directly correlate with the respective order parameters~\cite{Hu2017pre, Wetzel2017pre}.
    Besides structure recognition, NNs can be trained to predict macroscopic physics quantities such as the total energy, and microscopic quantities such as charge density and magnetization locally for each atom~\cite{Pasini2020jpcm}. NNs are also used to learn interatomic potentials, for example. They have been trained to generate effective many-body potentials from \textit{ab-initio} data~\cite{ZhangL2018prl} and were successfully applied to construct precise phase diagrams of water in molecular-dynamics (MD) simulations over a large range of temperatures and pressures~\cite{Zhang2021arxiv}. 
    Another ML potential, ANI-Al, was trained to obtain quantum-level accuracy and has been successfully combined with MD simulations to study shock physics in metals~\cite{Smith2021natcomm}. Training ML surrogate models is also becoming a useful technique to bridge different length and time scales in computer simulations, see~\cite{Diaw2020pre}, for example.

    Finally, NNs have been applied in the field of polymer model simulations to study transition between coil and globule structures and recognize Mackay--Anti-Mackay structures, for example~\cite{Wei2017pre,Xu2019pre}. Transitions between such crystalline structures in the solid phase are notoriously hard to simulate~\cite{Schnabel2009cpl} and advanced generalized ensemble methods have been developed to do so in the past~\cite{Schnabel2011jcp,Schnabel2009jcp}. These studies emphasize the benefit of knowing the conformational state of a system during the simulation and ML could contribute valuable information if structures can be reliably recognized. In this paper we will provide more evidence that this can indeed be achieved by employing NNs
    in the supervised recognition of low-energy configurations of polymers absorbed to a substrate (Sec.~\ref{sec:recog_nn}). Furthermore, we will show how unsupervised ML method can be employed if no previous knowledge of structural phases of a model is available beforehand (Sec.~\ref{sec:dim_reduction}). Finally, we will determine boundaries between phases in the model parameter space by training NNs in a conventional way, but also by applying the more recently developed confusion method (Sec.~\ref{sec:ml_boundaries}).   

\section{Model and Observed Structural Phases} \label{sec:model}

To model the nanotube--polymer composite we use a coarse-grained bead--spring description for the polymer~\cite{Stillinger1993pre} and an attractive interaction between the monomers and the one-dimensional, continuous string that is derived from a Lennard-Jones potential~\cite{Vogel2010prl,Vogel2010pp,Vogel2011cpc}. The latter contains two parameters, the effective thickness of the string, $\sigma_f$, and its attraction strength, $\epsilon_f$:
\begin{equation}
    V(r_\perp)\propto\epsilon_f\left(\frac{63}{64}\frac{\sigma_f^{12}}{r_\perp^{11}}-
    \frac{3}{2}\frac{\sigma_f^{6}}{r_\perp^{5}}\right)
    \label{eq:model}
\end{equation}
where $r_\perp$ is the perpendicular distance between a mono\-mer and the string.
The interaction between nonbonded monomers is described by a standard Lennard-Jones potential and there is a weak bending stiffness for consecutive mono\-mer--monomer bonds, as often employed in bead--spring polymer models~\cite{Stillinger1993pre,Bachmann2005pre,Schnabel2007prl}. For a more detailed discussion of different approaches to model a thin, cylindrical substrate, see~\cite{Vogel2015jcp}.

\begin{figure}
    \centering
    \includegraphics[width=\columnwidth]{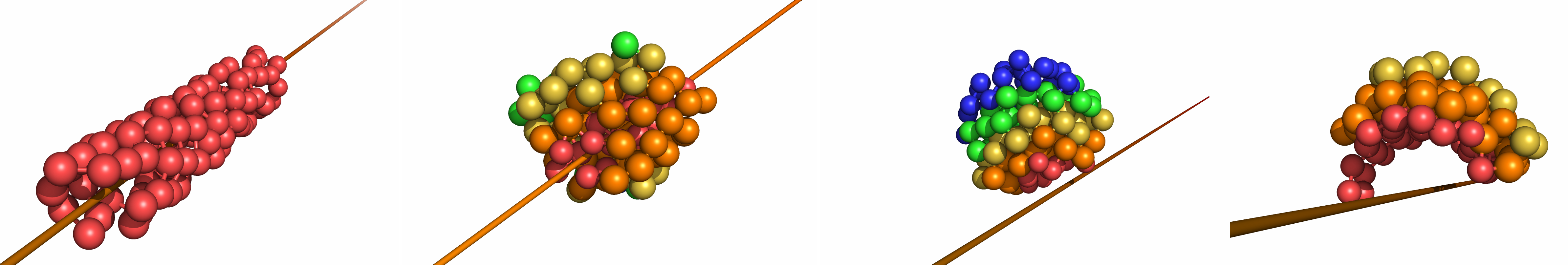}
    \caption{Examples of low-temperature configurations in phases B, Gi, Ge, and C (from left to right). Different monomer colors indicate their distance from the string.}
    \label{fig:phases}
\end{figure}

Depending on the parameter set $\{\sigma_f,\epsilon_f\}$ in Eqn.~(\ref{eq:model}) low-energy structures take qualitatively different shapes that can be grouped into structural classes or phases. In our previous work we distinguished between four such phases and labeled them Ge, Gi, C, and B; see Fig.~\ref{fig:phases} for visualisations. Structures like the ones we find in the Gi, Ge, and C regions have been found and imaged in experimental studies before, in particular ``clam-shell'' (C) polymer nanodroplets have been emphasized~\cite[Fig. 4]{Tran2008nl}. Globular configurations in phase Ge are similar to structures seen during dewetting of polymers on the surface of carbon nanotubes (CNTs)~\cite[Fig. 2]{Tran2008nl} while Gi configurations for large values of $\epsilon_f$ show similarities with ``barrel-type'' nanodroplets~[\textit{ibid}]. Note that pure monolayer barrel structures (B) can be mapped onto different types of CNTs~\cite{Vogel2013ccp,Vogel2011pp}. In fact, we found that region B contains sub-phases with different chiralities corresponding to those found in CNTs~\cite{Vogel2010pp}.\footnote{While those sub-phases should be able to be recognized by appropriately trained neural networks, we will not emphasize those any further in this paper.}

\section{Machine Learning of Structural Phases} \label{sec:ml_small}

In the following we investigate different supervised and unsupervised machine learning (ML) methods for structure recognition. In ML one typically desires large datasets to reliably train a robust model. However, in the research presented here the data is intrinsically hard to generate since we are analysing states that dominate canonical ensembles at very low temperatures. We use Wang--Landau (WL) sampling~\cite{Wang2001prl} to produce these low-energy configurations. Even though WL reliably finds these states, we face the challenge to have to collect many, very different and ideally uncorrelated low-energy configurations for all parameter values $\{\sigma_f,\epsilon_f\}$ (see Eq.~\ref{eq:model}). In an extreme approach and as a proof of concept, we here only record one configuration every time the WL walker explores a low-energy valley and then wait for the walker to move to regions in the phase space corresponding to high temperatures before collecting data again at low energies. Admittedly, such a strategy is computationally expensive and even though applied to generate the dataset analysed in this section, it might not be necessary in that extreme way (see a discussion below in Sec.~\ref{sec:ml_boundaries}).

In all our simulations, a polymer configuration is represented by the three 
spatial coordinates of $100$ monomers. We use these 300 coordinates (either in raw format or preprocessed, see below) as the feature set for the machine learning algorithms. Although it is possible to utilize an engineered feature set based on our physical intuition of the system (for example by including macroscopic physical observables like the radius of gyration, end-to-end distance, energy, etc.), avoiding such engineered features leaves the machine learning algorithm unbiased and free of any preconceived notions.

\subsection{Supervised Learning: Structure Recognition with Neural Networks}
\label{sec:recog_nn}

\begin{figure}[t]
    \includegraphics[width=\columnwidth]{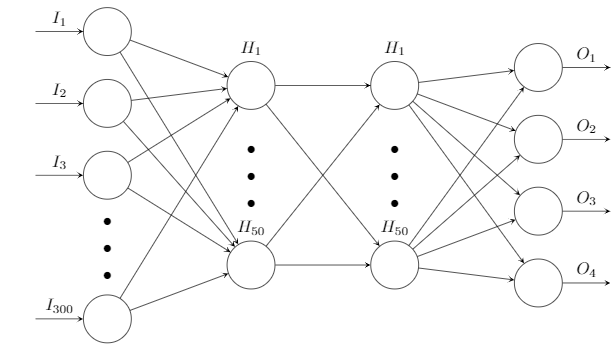}
    \caption{Schematic of the neural network with 300 input neurons, two hidden layers, an four output neurons.}
    \label{fig:nn_sketch1}
\end{figure}

\begin{figure*}
    \includegraphics[width=.95\textwidth]{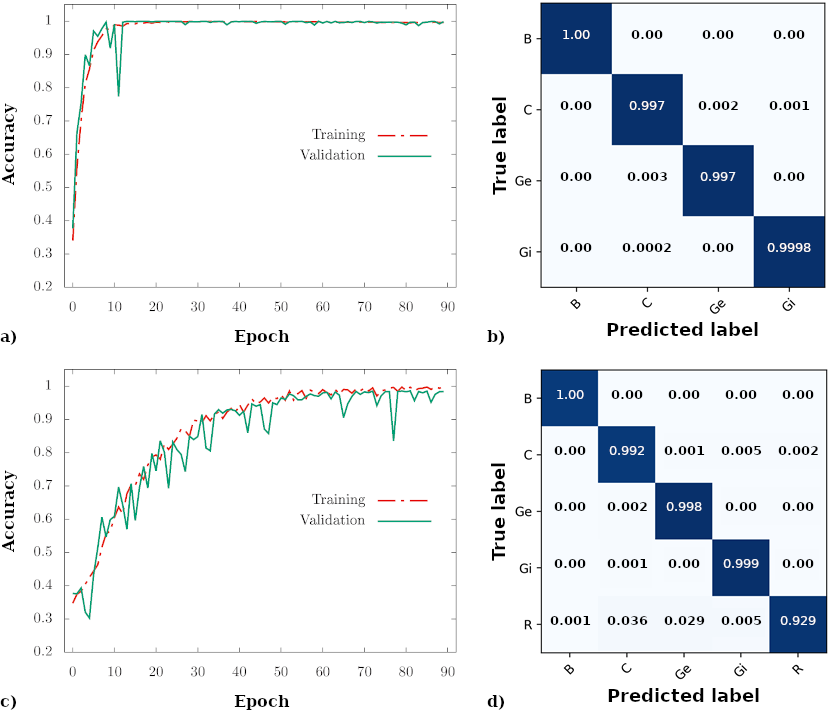}
    \caption{Training and validation accuracy (left) and normalized confusion matrices (right) during supervised structure recognition training on the the neural network. Top row: Only the four low-energy phases (B, C, Ge, Gi; where polymers are adsorbed on string) are used for training and recognition. Bottom row: Data includes high-temperature, random coil configurations (R; where polymers are desorbed from string).}
    \label{fig:nn_results1}
\end{figure*}

The neural network (NN) is set up with an input layer of 300 neurons, two hidden layers of 50 neurons each and one output layer of four neurons, see Fig.~\ref{fig:nn_sketch1}. The dataset consists of about 3000 configurations or samples for each polymer type. Two thirds of the dataset are allocated for training, while the remaining data is used for testing. The rectified linear unit (ReLU) activation function is used for the hidden layers and the softmax function for the output layer. We use the Nadam optimization algorithm over ninety epochs for training. Finally, to mitigate overfitting we employ L2 kernel regularization. The results of the NN classification are shown in Fig.~\ref{fig:nn_results1} where we plot the confusion matrices and the learning curves.

We start by running an analysis with the adsorbed polymer configuration types B, C, Ge, Gi (see Fig.~\ref{fig:phases}) as part of the dataset. We preprocess the data by spatially shifting the monomers in the $z$ direction such that the center of mass of each polymer lies on the $xy$ plane. Figure~\ref{fig:nn_results1}\,(a) shows the training and validation accuracy measured at each epoch during the NN training. Both the training accuracy and the validation accuracy rapidly converge to a steady value within 20 epochs. In addition, the training and validation curves are quite close to each other and therefore show no noticeable sign of overfitting. Figure~\ref{fig:nn_results1}\,(b) shows the confusion matrix obtained for the validation set, normalized by the number of elements in each class. We see that the off-diagonal elements are zero, except for a very few misclassifications of C-type polymers, yielding an almost 100\% overall validation accuracy.


Since the neural network was able to reliably identify all adsorbed polymer structures we also included high-temperature, random-coil polymer structures (``R'') not adsorbed to the string as another type and added an output neuron accordingly. Figures~\ref{fig:nn_results1}\,(c) and~(d) show the corresponding accuracy curves and the confusion matrix. The training and validation accuracies again converge to 1.0, although the convergence rate is slower compared to the previous case. Again, the curves do not show evidence of noticeable overfitting. The slightly increased presence of off-diagonal entries in the ``R'' row of the confusion matrix indicates a somewhat higher tendency for random coil configurations to be misclassified as other polymer types. However, this is expected as those polymers are random configurations that could, in fact, loosely resemble any of the other classes by chance. 

\subsection{Unsupervised Learning: Dimensionality Reduction Methods} \label{sec:dim_reduction}

Dimensionality reduction methods refer to a class of unsupervised machine-learning techniques
that map data from an original, high-dimensional space to a lower-dimensional space while ideally
preserving some of the salient properties of the data.
In the context of thermodynamic phase classification, for example, such low-dimensional representations of
the configuration space have been used to facilitate the visual identification of 
distinct phases~\cite{Wang2016prb, Wang2017prb, Wetzel2017pre}
and to provide insight into the relationship between important features and order parameters
of complex systems~\cite{Wetzel2017pre, Hu2017pre, Xu2019pre}.

Principal component analysis (PCA)~\cite{Pearson1901} is a linear dimensionality reduction
technique which identifies a set of mutually orthogonal unit vectors in a given feature space. These vectors are ordered according to the variance of the data in the corresponding directions, such that the first unit vector indicates the direction of greatest variance in the data. This direction is then called the first principal
component, the one with the second highest variance the second principal component, and so on. 
The principal components are the eigenvectors of the covariance matrix of the data, and hence can be determined by the eigendecomposition of that matrix or the singular value decomposition of the data matrix. The original configurations are then projected into a space spanned by the first $m$ principal components to obtain the desired lower $m$-dimensional representations. For some spin systems, the principal components have been shown to recover the physical order parameters for phase transitions~\cite{Hu2017pre, Wetzel2017pre}.\newpage

In addition to PCA, we apply a number of other non-linear dimensionality reduction methods,
namely, multidimensional scaling (MDS)~\cite{mds}, $t$-distributed stochastic neighbor 
embedding ($t$-SNE)~\cite{tsne}, Isomap~\cite{isomap}, and diffusion map~\cite{diffusionmap}. 
Note that Isomap becomes equivalent to PCA as the neighborhood size approaches the sample size. Therefore, we limited the neighborhood size to 20 for this demonstration, but also confirmed that changing this number will not change the qualitative results.
In general, such non-linear methods identify lower-dimensional manifolds embedded within
the higher-dimensional feature space, in which similar data points are clustered together.
Typically, manifold\break\newpage\noindent learning methods can capture nonlinear relationships within the data 
that cannot be captured through principal component analysis.

\subsubsection{Data Pre-processing}

When employing unsupervised learning methods the data typically has to be prepared in some way to obtain meaningful results. In the raw data the polymer is adsorbed at the string at an arbitrary position, while the string is always located at the $z$-axis in Cartesian coordinates. We here utilize different scaling and coordinate transformation methods to potentially make the features of the polymers more comparable for the machine. A common method in machine learning, referred to as ``standard scaling''~\cite{standard_ML_text} aims at bringing all features (in our case, monomer coordinates) onto the same length scale by subtracting the mean of all data from each feature and individually scaling each feature to unit variance. Two other ways that do not alter the overall shape of the polymer are translations along the string such that the $z$-component of the center of mass is zero for all polymers, eliminating arbitrary shifts in spatial position, and translations of the overall center of mass to the coordinate origin, normalizing the position of the polymers across the simulated examples. While the first aims at recognizing the general position of the polymer with respect to the string, the latter is aimed at identifying the internal structure of globular polymers. Ge and Gi type configurations, for example, have a similar surface shape, but differ in relative position to the string and their internal crystalline structure. Finally, to help the machine recognize structural rather than size differences across all polymer types, we scaled all polymers with respect to their  radius of \hbox{gyration~$R_g$.}

\subsubsection{Results of unsupervised learning}

\begin{figure*}
    \centering
    \includegraphics[width=\textwidth]{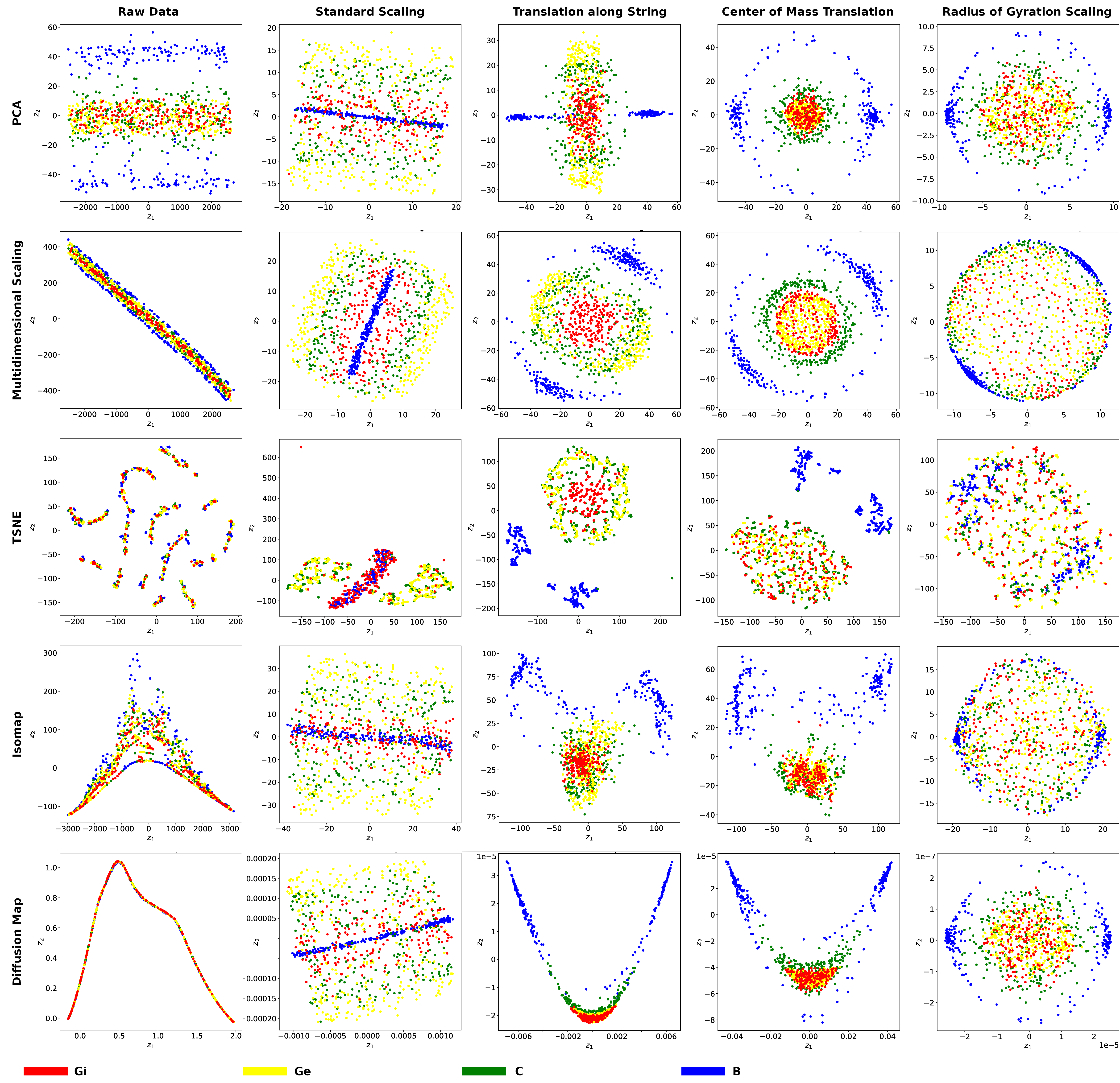}
    \caption{Two-dimensional projections of the configuration space obtained using various dimensionality reduction techniques, namely, principal component analysis (PCA), multidimensional scaling (MDS), $t$-distributed stochastic neighbor embedding ($t$-SNE), Isomap, and diffusion map. Different columns represent different 
    data preprocessing techniques applied: (from left to right) raw data without preprocesssing, data standardized by subtracting the mean and scaling to unit variance, data processed by subtracting the $z$ component of the center of mass, data processed by subtracting all three Cartesian components of the center of mass, and data scaled by the radius of gyration after subtracting the center of mass.
    }
    \label{fig:matrix}
\end{figure*}

In Fig.~\ref{fig:matrix} we present the two-dimensional representations of the configuration space obtained from
the dimensionality reduction methods mentioned above. Different columns show results after different data preprocessing steps employed (raw data without preprocesssing, standard scaling, subtracting the $z$-component of the center of mass, subtracting
all three components of the center of mass, and data scaled by $R_g$ after subtracting the center of mass). 
Even without any preprocessing (leftmost column), for example, we observe that PCA can reasonably well distinguish barrel-like (B) conformations from all others. However, none of the methods can distinguish Gi, Ge, and C conformations without preprocessing. This is presumably because the $z$ coordinates of the
configurations have much higher variance than the $x$ and $y$ coordinates, as the system shows translational
invariance in the $z$ direction. The poor performance of ML algorithms due to different features having different
scales is a common problem in machine learning. This issue can be alleviated with appropriate feature scaling
techniques. Here we first test standard scaling. As the second column of
Fig.~\ref{fig:matrix} shows, this approach does improve the performance of the algorithms (particularly that of
MDS), as a clearer separation of Gi, Ge, and C conformations can be observed. However, it is important
to note that since the scaling is performed independently on individual features, these coordinate transformations
lead to non-physical deformations in the polymer configurations.

A more physically intuitive scaling approach is to subtract the center of mass, which would reduce the variance
of the coordinates due to the drifting of polymers in arbitrary directions. In particular, polymers have the freedom
to drift along the substrate in $z$-direction. Therefore one would 
expect noticeable improvements in the results just by subtracting the $z$ component of the center of mass alone.
As the third column of Fig.~\ref{fig:matrix} shows, we indeed observe improved performance in most algorithms in
terms of separating previously overlapping phases observed in the analysis using raw data.

The fourth column shows
the results obtained by subtracting the whole center of mass. For some algorithms
(particularly  MDS), subtracting all three components of the center of mass further improves phase separation.
The rightmost column in Fig.~\ref{fig:matrix} shows the results obtained with all coordinates furthermore normalized by scaling with the radius of gyration $R_g$. However,
we observe that Gi, Ge, and C phases are no longer distinguishable. This indicates that the length scale of polymers
is a particularly important feature for distinguishing different polymer states.

In summary, we find that identifying barrel-type (B) configurations can be accomplished by all methods with suitable preprocessing steps. Telling all other structures apart is more challenging and no single scheme is able to do so alone.\footnote{We note that it might be possible to do so with a reduction to a 3-dimensional space though.} That said, 
we note that the MDS method trained with data preprocessed by subtracting all three components of the
center of mass seems to give the best, single overall performance, particularly since both B and C conformations are grouped into isolated
clusters spatially separated from other states. Still, an overlap between Gi and Ge phases can be observed in this case since both phases differ mostly by the relative location with respect to the substrate and not in overall shape. To observe a separation between Gi and Ge structures one would need to use another procedure, like MDS or PCA with a translational normalization along the $z$-direction only. The general finding that no one scaling approach and reduction-method combination can clearly separate all phases present in our system is probably true for other complex polymer systems as well.\enlargethispage{-\baselineskip} Depending on symmetry and specific structures, different data preprocessing and scaling methods might always have to be chosen to match all physical \hbox{properties}.

\section{Identifying Structural Transitions with Neural Networks} \label{sec:ml_boundaries}

In this section we study the applicability of neural networks (NNs) to not only recognize different structures but to detect transitions points between them. While we have discussed above the desire to use large datasets for NN training in general, much more training data is potentially needed for such an endeavor since structural differences could be much more subtle between polymers close to each other in parameter space, compared to above (Sec.~\ref{sec:ml_small}) where structures are more fundamentally different from each other. To enrich our datasets we therefore apply a strategy in the spirit of oversampling augmentation~\cite{shorten2019jbd} where we allow to record up to 100 slightly modified configurations every time the WL walker explores a low-energy region. After reaching that number, the walker has to completely ``warm up'' again, that is move to energies encountered well inside the random-coil phase. To ensure the data in each such batch is not effectively identical but to some degree still uncorrelated we enforce a minimum energy difference $\Delta E$ between two consecutive configurations that are added to the dataset. 

\subsection{Conventional, supervised approach} \label{sec:supervised_approach}

\begin{figure}[t]
    \centering
    \includegraphics[width=\columnwidth]{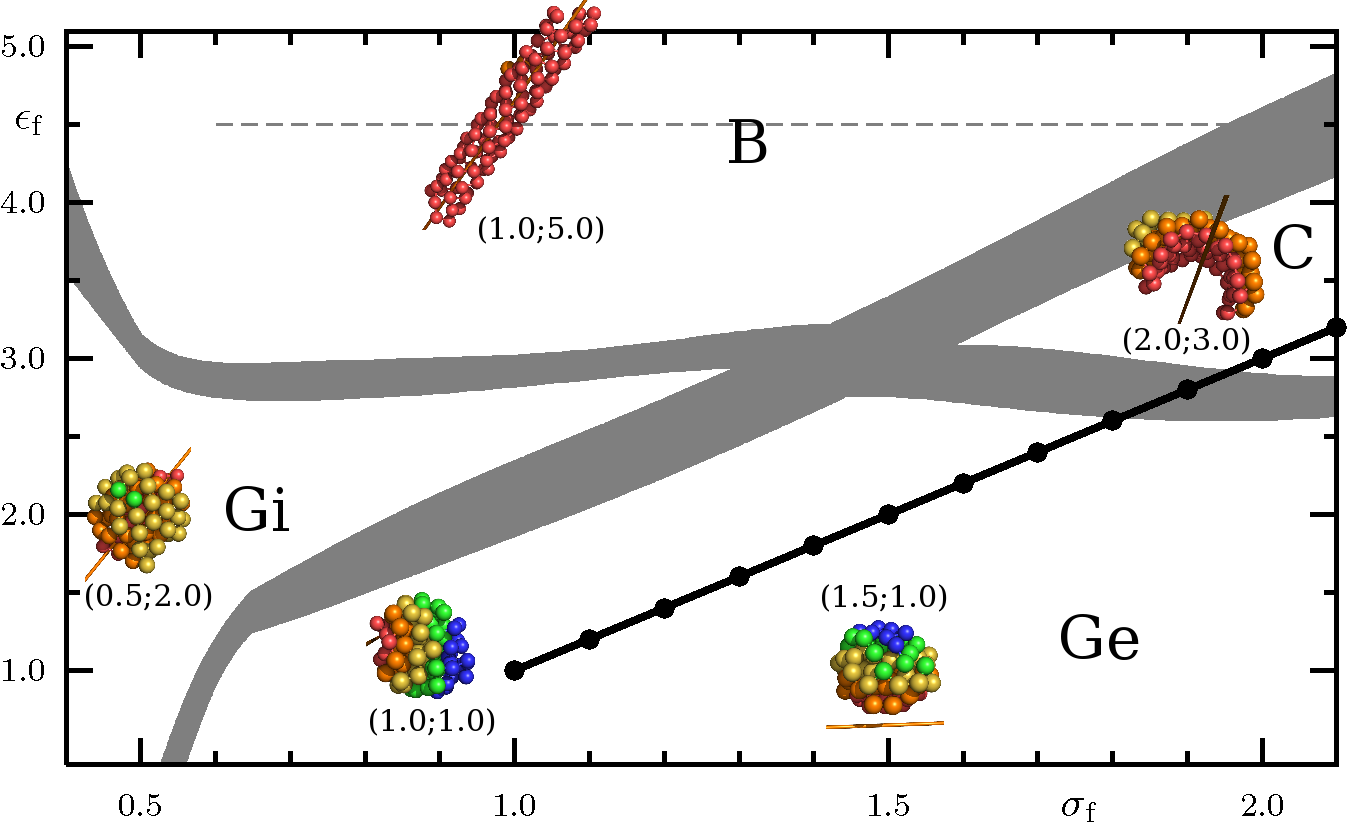}
    \caption{Configurational phase diagram in model-param\-eter space of low-temperature polymer configurations adsorbed to a thin string. Previously determined transition regions are shaded in gray. We analyse structures at all points indicated along the diagonal line from $\lambda=\{\sigma_f, \epsilon_f\}=\{1.0,1.0\}$ to $\lambda=\{2.1,3.2\}$.}
    \label{fig:phase_diagram}
\end{figure}

In previous research we had to rely on human intuition to define structural classes and suitable observables or order parameters to find the boundaries in parameter space between them. The structural phase diagram for low-energy states~\cite{Vogel2010prl} (see Fig.~\ref{fig:phase_diagram} for a reduced version) was hence developed upon the \textit{ad-hoc} introduction of an asymmetry parameter, for example, to locate the crossing from phase Gi to B. Such practice inevitably introduces a bias based on the human perception of structure. It is therefore, in principle, hard to judge whether or not we identified the most relevant structural features. A less biased approach that currently gets increasing attention is the use of machine learning methods to identify crossing or phase transition points between structural or thermodynamic phases~\cite{Wei2017pre,Carrasquilla2017np,rem2019nat,Walters2019pre,zhang2019pre,Munoz2020jsm}. We here use neural networks that we train with data which can be clearly assigned to different structural classes and have them analyse polymer configurations in regions of the parameter space where such a classification is less defined.

\begin{figure*}[ht]
    \centering
    \includegraphics[width=\columnwidth]{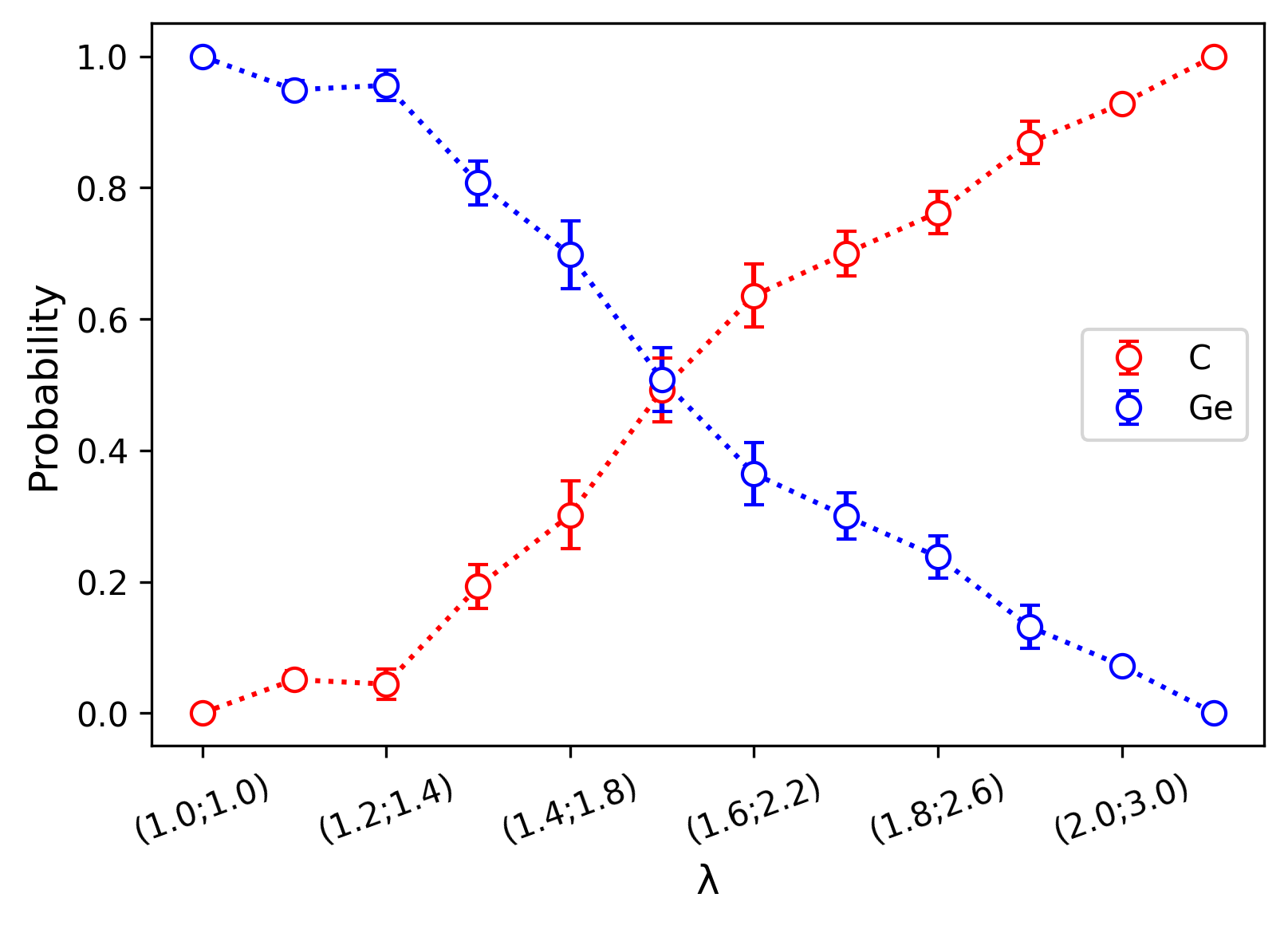}
    \includegraphics[width=\columnwidth]{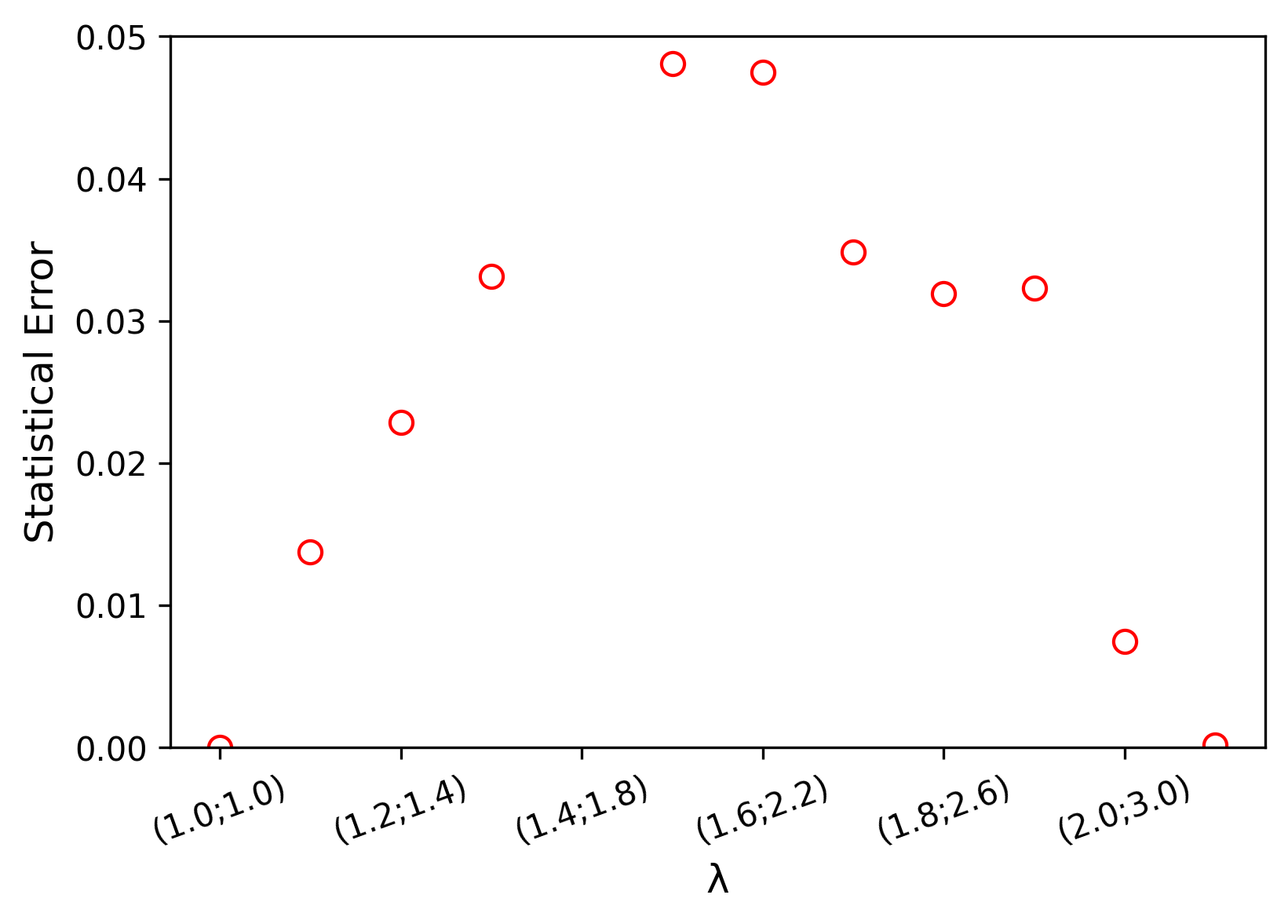}
    \caption{Left: probabilities of predicting C and Ge polymers at different points in parameter space after training the NN with data at $\lambda= \{\sigma_f, \epsilon_f\} =\{1.0,1.0\}$ and $\lambda=\{2.1,3.2\}$ (outermost points on each side). Right: The corresponding statistical error from different, independent predictions. The error is largest around the transition region.}
    \label{fig:sec4_GeC}
\end{figure*}

Specifically, we investigate the transition between globular polymers absorbed to the string (Ge) and clam-shell structures surrounding the string (C), two of the phases that were particularly hard to distinguish with the unsupervised methods discussed above. The NN is set up with the same parameters as above (cf. Fig.~\ref{fig:nn_sketch1}) with the difference that only two nodes are specified for the output layer. The network is then trained with configurations at $\lambda = \{\sigma_f, \epsilon_f\} = \{1.0,1.0\}$ and $\lambda=\{2.1,3.2\}$, which clearly belong to the Ge and C classes, respectively. We use the such trained network to analyse configurations at ten other parameter values in between those points (see black, diagonal line in Fig.~\ref{fig:phase_diagram}) and predict their belonging to either class. When plotting the corresponding probabilities, as shown in Fig.~\ref{fig:sec4_GeC} (left), we see a ``crossing'' of the probability curves. As one would expect, the corresponding error bars are largest around the phase intersection and decrease towards the outermost points, see Fig.~\ref{fig:sec4_GeC} (right). That is, the uncertainty of the network in classifying polymer configuration is maximal around the transition from one phase to another. We assess uncertainties of the trained NN models via different methods including cross-validation~\cite{Goodfellow2016} and query-by-committee~\cite{lakshminarayanan2017}. In cross-validation, a subset of the whole dataset is held out for testing while the remaining data would be used for training. The process repeats with different held-out testing sets, resulting in a group of NN models that can be used for the estimation of statistical errors. We performed 10-fold cross-validation but it seemed to underestimate the real error of the model. 
It could be because the 10 resulting models are not statistically independent: each of them are trained using training datasets that overlap with each other by 80\%. When these highly-correlated models are applied to make predictions on out-of-sample data (structures between $\lambda  = \{1.0, 1.0\}$ and $\lambda  = \{2.1, 3.2\}$), they result in a small distribution (variance) around the mean, but the mean prediction might have a high discrepancy (bias) compared to the reference.
Hence we report errors from query-by-committee: the whole dataset is divided into 10 subsets, within each subset 70\% of the data were used for training and 30\% of the data were used for testing. This results in 10 individually trained NN models that are truly independent and not correlated. The error bars we show in Fig.~\ref{fig:sec4_GeC} therefore indicate the statistical error from multiple runs with NNs which were individually trained with independent data and also analysing different datasets. That way we capture both epistemic and aleatoric uncertainties. 

\subsection{Unsupervised: The Confusion Method}

A neural network is inherently a supervised learning method and requires a
dataset with preassigned labels for the adjustment of weights in the training phase.
However, in certain cases it may be difficult, or even impossible, to know the 
correct assignment of labels beforehand.
For the case of phase classification, 
one can circumvent this issue by identifying a 
window within which a given transition occurs
and labeling the configurations outside this window based on the corresponding
phase labels, as we did above in Sec.~\ref{sec:supervised_approach}.
In particular for finite systems that do not
naturally scale up to the thermodynamic limit, though, it can be challenging to reliably locate
the exact point of transition this way. 

The confusion method~\cite{Nieuwenburg2017np} provides an alternative. It not only eliminates the need for prior assignment of labels,
but also results in a clearer and more precise estimate of the transition point. This method as well is a neural-network based, 
but semi-supervised approach for detecting phase transitions and relies on purposeful mislabeling of the data.
Let $\lambda$ denote a model parameter or a thermodynamic observable
such as the temperature or the average energy.
Assume that there exists a critical point $\lambda_c$ at which a transition
from a phase X to a phase Y occurs.
When applying the confusion scheme, one first identifies a 
window $[\lambda_a, \lambda_b]$ within which the transition is likely to occur.
Then a potential transition point $\lambda_c^\prime \in [\lambda_a, \lambda_b]$ is proposed, 
and the label ``0'' (denoting phase X) is assigned to all configurations below $\lambda_c^\prime$, 
and the label ``1'' (denoting phase Y) to all configurations above $\lambda_c^\prime$. 
A neural network is then trained with this label assignment and the classification accuracy  
$P(\lambda_c^\prime)$ obtained for a test set is recorded. 
This process is repeated by systematically varying $\lambda_c^\prime$ 
from $\lambda_a$ to $\lambda_b$.
The resulting curve $P(\lambda_c^\prime)$ then
yields a characteristic ``W'' shape, with the middle peak occurring
at $\lambda_c^\prime = \lambda_c$~\cite{Nieuwenburg2017np}.

This W-shaped profile of $P(\lambda_c^\prime)$ can be understood as follows.
For $\lambda_c^\prime = \lambda_a$, all configurations are labeled ``1'',
and the neural network correctly predicts the assigned label for all samples,
achieving 100\% accuracy.
Similarly, the network performs with 100\% accuracy for 
$\lambda_c^\prime = \lambda_b$ as all the configurations are labeled ``0''.
For $\lambda_c^\prime=\lambda_c$, the assigned labels for all samples
exactly match the true phase labels
and, in principle, the NN can again achieve perfect accuracy.
For other values of $\lambda_c^\prime$, the NN sees a discrepancy between the assigned labels
and the true phase labels as identified by the patterns in data.
Due to this confusion, the NN learns to predict the majority label.
Ultimately this yields the characteristic W shape of $P(\lambda_c^\prime)$.
Note that in practice this shape will likely be distorted due to finite-size effects and imperfections in the training process. 

\begin{figure}[t]
    \includegraphics[width=\columnwidth]{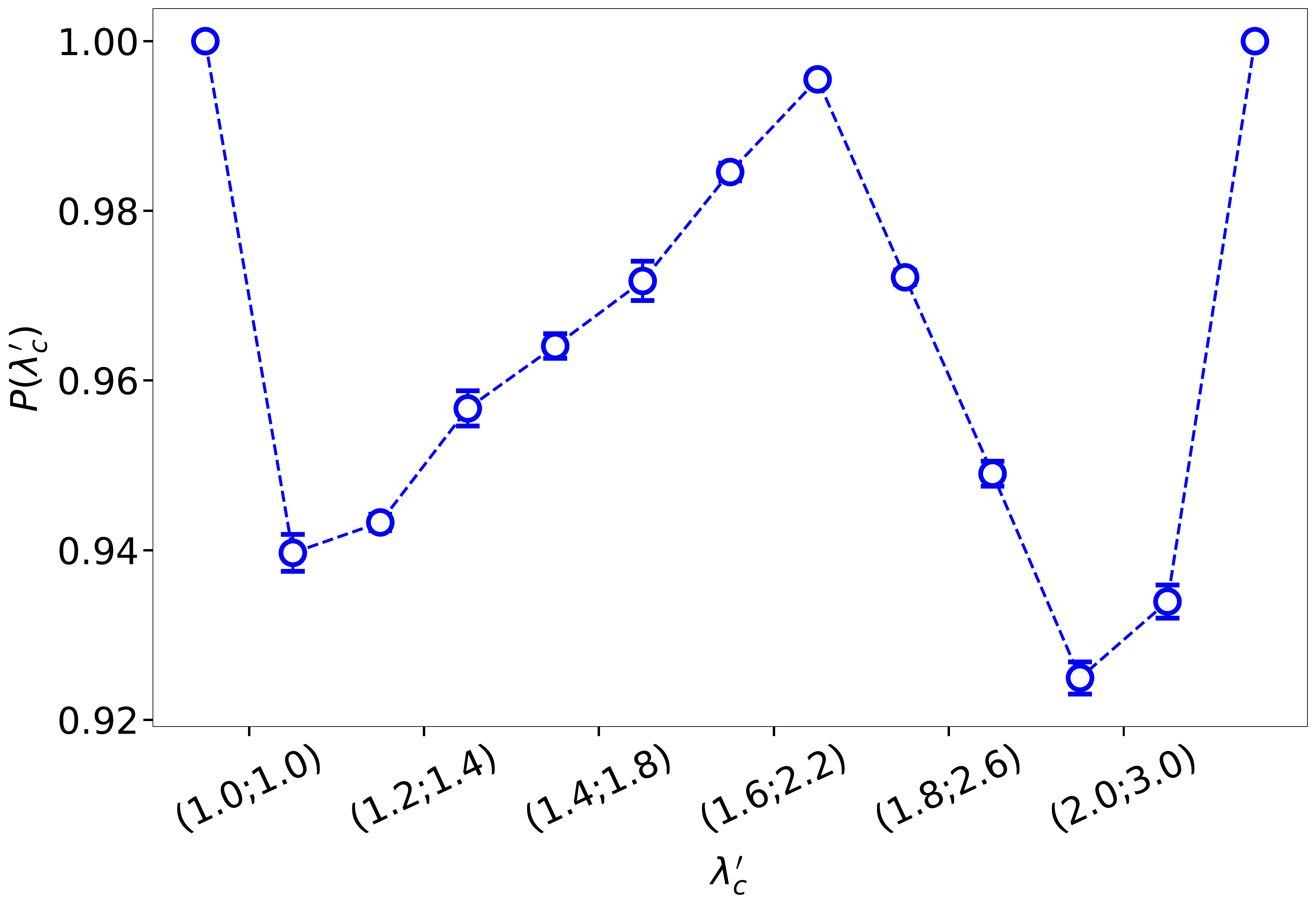}
    \caption{The test accuracy $P(\lambda_c^\prime)$ as a function of the proposed transition
    point $\lambda_c^\prime$.}
    \label{fig:confusion_method}
\end{figure}

We apply the confusion method to investigate the transition from the Ge phase 
to the C phase along the same straight path through the phase diagram as above and indicated in Fig.~\ref{fig:phase_diagram}.
For the neural network we adopt a feed-forward architecture with a single hidden layer of 40 neurons. We use the same data as above and 70\% of the configurations were randomly selected for training while the remaining samples were used for testing.
For error estimation, the confusion scheme is repeated 10 times, each time with a different 
random selection of training and testing samples.
Figure~\ref{fig:confusion_method} shows the test accuracy $P(\lambda_c^\prime)$ as a function of the proposed transition point $\lambda_c^\prime$. The curve follows the characteristic W shape discussed above, confirming the expected transition from the Ge phase to the C phase.
The middle peak indicates the location of the transition point $\lambda_c^\prime = \lambda_c$.

We emphasize that the evidence for a transition provided by the confusion method is more compelling than that provided by the conventional neural network based
approach discussed above in Sec.~\ref{sec:supervised_approach}.
The conventional approach may falsely indicate the presence of a transition within the
window $[\lambda_a, \lambda_b]$ even when all samples belong to the same phase, because also in the absence of a transition configurations may undergo slight structural changes as the parameter $\lambda$ is varied between $[\lambda_a, \lambda_b]$.
More specifically, one assumes that the transition occurs within a sub-window 
$[\lambda_1, \lambda_2]$, and assigns the label ``0'' to all configurations in the interval 
$[\lambda_a, \lambda_1]$, and the label ``1'' to all configurations in $[\lambda_2, \lambda_b]$.
The neural network may then detect the gradual structural changes in the configurations as a function of $\lambda$, and establish a decision boundary between $\lambda_1$ and $\lambda_2$
such that the predictions are consistent with the assigned labels. Consequently, the curves for the averaged output neuron values may still cross each other, giving the false indication of a transition.
In the confusion method however, the W-shape profile is only possible if there are abrupt, drastic changes in structure at a certain value of $\lambda$, reminiscent of a true transition.
In the absence of a transition, the middle peak in $P(\lambda_c^\prime)$
curve disappears, resulting in a ``V'' shape~\cite{Nieuwenburg2017np}.

\section{Summary} \label{sec:summary}

In this paper we explore the applicability of various machine learning methods to recognize structures and structural transitions in a model for polymer--nanotube composites. In particular, we investigate structures that have been observed experimentally where the polymer is adsorbed at the nanotube. The two main questions we address are whether and how we can identify those structure with machine learning and how to locate the transition regions between them.

For structure recognition we test various unsupervised dimensionality reduction methods like principal component analysis or multidimensional scaling that we combine with different ways to pre-process the data. The advantage of unsupervised methods is that no pre-labelling of structures is required, removing all potential human bias in structure classification. We find that while structure identification in principle is possible, no single method alone is capable of doing so. We found it particularly challenging to have the machine differentiate between globular structures where the polymer is fully wrapped around the substrate or just connects to the tube.
Aside from the unsupervised methods we also employed neural network methods that do require pre-labeled input. The network was able to reliably recognize all polymer structures after suitable training.

While it is probably uncontroversial to introduce different structural phases for polymer--nanotube composites, finding the exact boundary between those phases remains a challenge since it is in general not obvious what good order parameters are. We previously introduced such parameters \textit{ad-hoc}, but test here if a neural network could identify transitions between configurational phases, with and without training using configurations from the respective phases and without further human guidance or knowledge of pre-defined order parameters. This will be particularly useful since there is not a sharp, thermodynamic phase transition. We find that neural-network methods still indicate a transition, most notably the confusion method. However, since such structural transitions of finite systems potentially happen in different steps and over a broader region in parameter space, different machine learning methods or neural networks might pick up different steps in this transition at slightly different parameter values. In that sense, results for the crossing point shown in Figs.~\ref{fig:sec4_GeC} and~\ref{fig:confusion_method} are not necessarily contradicting, when keeping also in mind that data has to be binned for the confusion method, leaving a corresponding uncertainty in the exact position of the crossing. That said, we also note
that the traditional method of training the network with labeled configurations from both phases has to be used with care since it can potentially detect a crossing even if there was no phase transition. The main advantage of the confusion method here is therefore that it provides evidence for a transition between Ge and C. Otherwise, the shape of the detection accuracy graph would be a ``V''-shape rather than a ``W''-shape.

Overall, we confirm that defining structural transitions in our system is reasonable, in principle. We also conclude, though, that we might not have been successful initially~\cite{Vogel2010prl} in finding the best order parameter for all transitions, in particular for the crossing between "Ge" and "C" structures, as evidenced by the results shown in Figs.~\ref{fig:phase_diagram}--\ref{fig:confusion_method}. In that sense, the machine learning methods applied here can be a valuable complement to more conventional methods of detecting structural transitions used earlier, as they remove the necessity of identifying or defining explicit order parameters beforehand and therefore provide a potentially less biased approach to structure recognition and classification.
    
\section*{Acknowledgements} \label{sec:acknowledgements}

All data was produced on the University of North Georgia's Pando computing cluster. Y. W. Li acknowledges support from U.S. Department of Energy, Office of Science, Office of Basic Energy Sciences, under Award Number DE-SC0022311. LA-UR-21-30102. 

\bibliography{main}

\end{document}